\documentclass[%
 aip,
 amsmath,amssymb,
 reprint,%
]{revtex4-1}

\usepackage{graphicx}
\usepackage{dcolumn}
\usepackage{bm}

\usepackage[utf8]{inputenc}
\usepackage[T1]{fontenc}
\usepackage{mathptmx}
\usepackage{etoolbox}
\usepackage{hyperref}

\makeatletter
\def\@email#1#2{%
 \endgroup
 \patchcmd{\titleblock@produce}
  {\frontmatter@RRAPformat}
  {\frontmatter@RRAPformat{\produce@RRAP{*#1\href{mailto:#2}{#2}}}\frontmatter@RRAPformat}
  {}{}
}%
\makeatother
\begin{document}

\preprint{AIP/123-QED}

\title[]{Probing Quantum Anomalous Hall Transport Under Microwave Irradiation Using a Topological Circulator}
\author{Athul Ashok}
 \altaffiliation[Current address: ]{Institute for Quantum Computing and Department of Physics and Astronomy, University of Waterloo, Waterloo, ON N2L 3G1, Canada.}
\author{Frank Jin}%
\affiliation{ 
Lawrence Livermore National Laboratory, Livermore, CA 94550, USA.
}%
\author{Nick Du}%
\affiliation{ 
Lawrence Livermore National Laboratory, Livermore, CA 94550, USA.
}%

\author{Luis A. Martinez}%
\affiliation{ 
Lawrence Livermore National Laboratory, Livermore, CA 94550, USA.
}%

\author{Jenny Zhou}%
\affiliation{ 
Lawrence Livermore National Laboratory, Livermore, CA 94550, USA.
}%

\author{Sean O'Kelley}%
\affiliation{ 
Lawrence Livermore National Laboratory, Livermore, CA 94550, USA.
}%

\author{Zachary J.-R. Espley}%
\affiliation{ 
Lawrence Livermore National Laboratory, Livermore, CA 94550, USA.
}%

\author{Gang Qiu}
\altaffiliation[Current address: ]{Department of Electrical and Computer Engineering, University of Minnesota, Minneapolis, MN 55455, USA}
\affiliation{Department of Electrical and Computer Engineering, University of California, Los Angeles, Los Angeles, CA 90095, USA.
}%

\author{Kang L. Wang}
\affiliation{Department of Electrical and Computer Engineering, University of California, Los Angeles, Los Angeles, CA 90095, USA.
}%

\author{Dong-Xia Qu}
 \email{qu2@llnl.gov}
\affiliation{ 
Lawrence Livermore National Laboratory, Livermore, CA 94550, USA.
}%

\date{October 4, 2026}

\begin{abstract}
Edge magnetoplasmons (EMPs) provide a platform for probing chiral charge dynamics and nonreciprocal microwave transport in topological quantum materials. However, detecting small perturbations to EMP propagation remains challenging because their signatures in conventional microwave scattering measurements can be weak. Here, we investigate microwave-photon-induced perturbations of EMP transport using a quantum anomalous Hall topological circulator. Pump--probe measurements reveal a strongly frequency-selective response: while microwave irradiation substantially modifies the EMP transmission at several pump frequencies, the transmission remains nearly unchanged at selected frequencies, including $4$ and $7~\mathrm{GHz}$. We further exploit the non-Hermitian mode hybridization of the coupled EMP--resonator system to probe the response near an exceptional point (EP). The pump-induced change in transmission magnitude near the EP is approximately twice that observed away from the EP, demonstrating an enhanced microwave response to perturbations. These results reveal the interplay among chiral EMP transport, microwave-photon-induced perturbations, and non-Hermitian dynamics, and demonstrate the potential of topological circulators as a promising platform for enhanced microwave spectroscopy and sensing.
\end{abstract}

\maketitle

\begin{figure} [t]
\centering
\includegraphics[width=0.85\columnwidth]{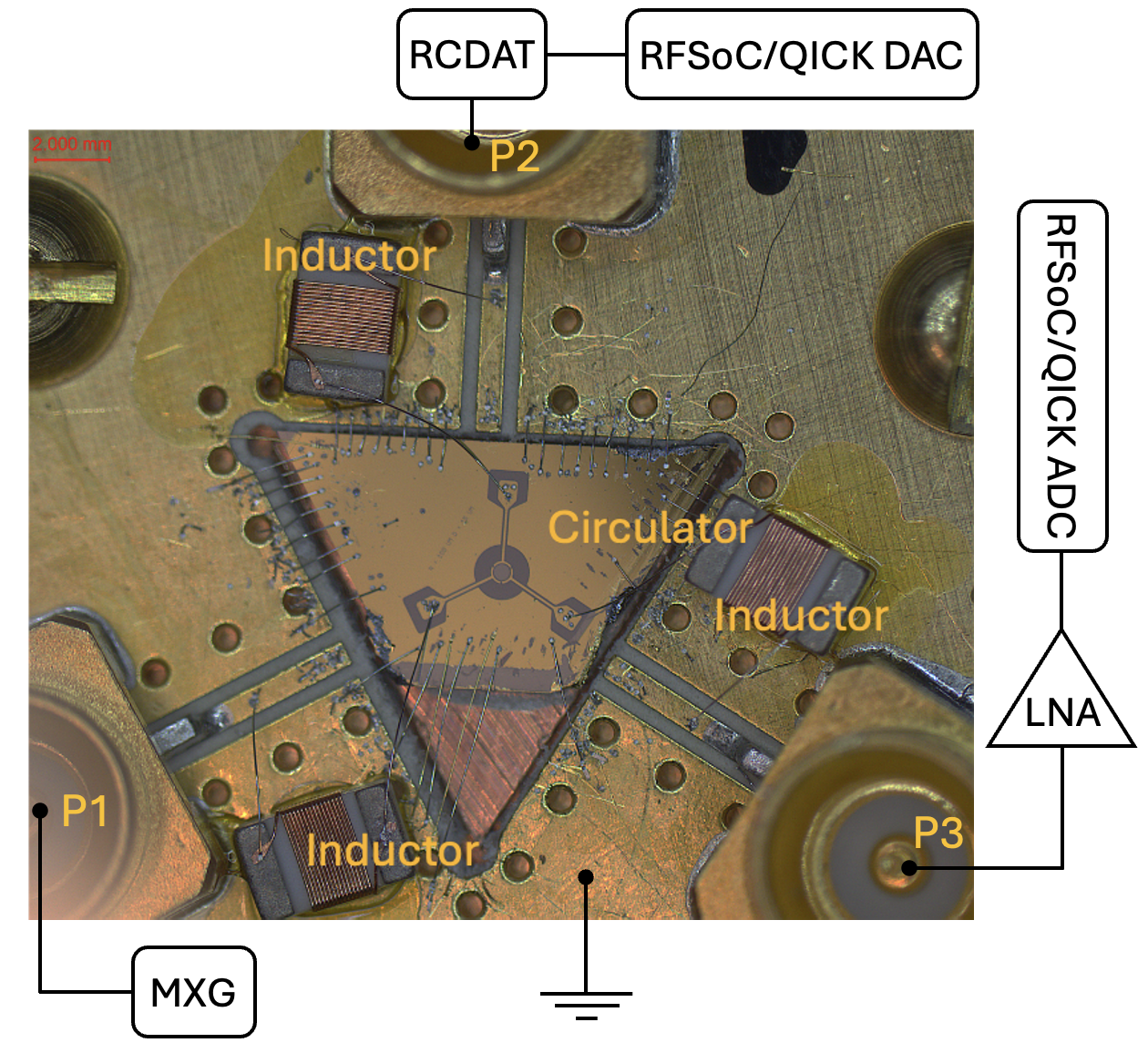}
\caption{\label{Figure1} Pump--probe measurement configuration for the topological circulator. A continuous-wave pump generated by the vector signal generator (Keysight N5182B MXG) is applied to Port 1 (P1). The Zynq UltraScale+ RFSoC board with the Quantum Instrumentation Control Kit (QICK) digital-to-analog converter (DAC) supplies the probe signal to Port 2 (P2) through a programmable attenuator (RCDAT), which sets the probe power incident on the device. The resulting transmission is collected at Port 3 (P3), amplified by a low-noise amplifier (LNA), and digitized by the RFSoC/QICK analog-to-digital converter (ADC).}
\vspace{-20pt}
\end{figure}

\begin{figure*}[!t]
\centering
\includegraphics[width=0.85\textwidth]{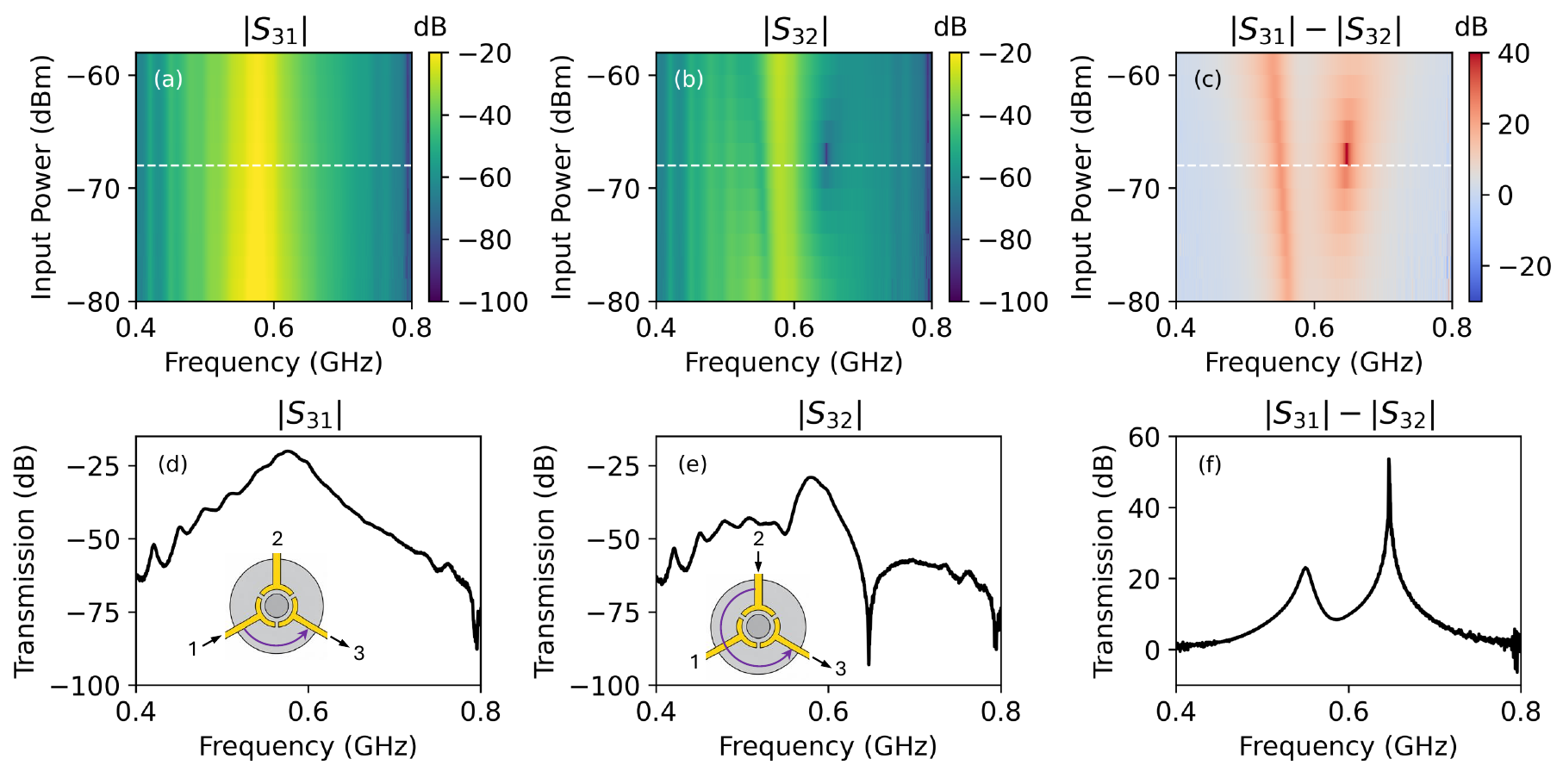}
\caption{{\bf Transmission response of the circulator as a function of input power.} (a)--(c) Power-dependent microwave transmission responses $|S_{31}|$, $|S_{32}|$, and $|S_{32}|-|S_{31}|$ for the $100\,\mathrm\mu m$-radius circulator coupled to LC resonators. The signal is excited from ports 1 and 2, and measured at port 3. (d)--(f) Frequency line cuts from panels (a), (b), and (c) at an input power of $P=-68 \ \mathrm{dBm}$. The insets in (d) and (e) show schematic device structures illustrating anticlockwise EMP transport from Port 1 to Port 3 and from Port 2 to Port 3, respectively. }\label{Figure2}
\end{figure*}
\section{Introduction}
Compact nonreciprocal microwave components are important for controlling signal flow and suppressing unwanted feedback in scalable cryogenic quantum systems.\cite{PhysRevX.5.041020, Mahoney2017,  PhysRevResearch.6.013081,Kumar2025Circulator, Martinez2026Circulators} Quantum anomalous Hall (QAH) materials provide a promising route to this functionality by combining magnetic order with the nontrivial electronic structure of a topological insulator.\cite{Chang2013, RMP2023} In a topological insulator, strong spin--orbit coupling gives rise to conducting boundary states with Dirac-like dispersion and spin-momentum locking \cite{RevModPhys.82.3045,Qu2010, Qu2012,Qu2018, Qu2024}. When time-reversal symmetry is broken by magnetic order, the system can enter the QAH state, characterized by an insulating bulk and chiral edge transport without the large external magnetic fields required in conventional quantum Hall systems. These unidirectional edge channels can couple directly to microwave fields, making QAH materials useful not only for compact nonreciprocal devices but also for probing the high-frequency dynamics and dissipation of topological edge transport \cite{PhysRevLett.124.076802,PhysRevResearch.6.013081}.

At microwave frequencies, the chiral edge channels support collective charge-density excitations known as edge magnetoplasmons (EMPs).\cite{Volkov1988EMP, PhysRevB.84.045314} An EMP corresponds to a propagating modulation of the edge charge density whose velocity and dispersion are determined by the edge-state conductivity, electrostatic environment, and the width of the one-dimensional edge state. Because the propagation is chiral, microwave signals coupled to the edge accumulate direction-dependent phase and attenuation, providing a natural mechanism for nonreciprocal transmission. \cite{Mahoney2017PRX} In a three-port geometry, capacitive coupling between microwave electrodes and the propagating EMP forms the basis of a topological microwave circulator. Previous measurements of QAH circulators have characterized the dispersion, drift velocity, and width of EMP modes and demonstrated substantial microwave nonreciprocity \cite{Mahoney2017, PhysRevResearch.6.013081,Martinez2026Circulators}.

The microwave response of the EMP is also sensitive to dissipative processes within the QAH material. Broadband microwave measurements have shown that coupling between the chiral edge state and charge puddles in the bulk modifies EMP propagation and introduces additional dissipation \cite{PhysRevResearch.6.013081, Roper2024Propagation}. Microwave excitation can further promote hopping between localized bulk states and ultimately contribute to breakdown of the QAH state \cite{Roper2025Breakdown}. Because the EMP propagates along the edge while remaining sensitive to dissipative bulk excitations through resistive and capacitive coupling, its microwave response provides a means of probing how applied microwave irradiation modifies the coupling between edge and bulk transport in the QAH system.

When the EMP mode is coupled to two LC resonators, the resulting hybrid system can be described by an effective non-Hermitian Hamiltonian because both modes experience loss and, in the circulator geometry, direction-dependent coupling. As the relative resonance frequencies, coupling strengths, and dissipation rates are varied, the coupling between the two LC resonators can become strongly asymmetric, and the system approaches a Hatano–Nelson-type exceptional point (EP) when one directional coupling is suppressed to zero while the reverse coupling remains finite.\cite{PhysRevLett.77.570} Near an EP, the system can exhibit an enhanced response to small perturbations because its eigenvalue becomes highly susceptible to changes in the underlying system parameters \cite{PhysRevLett.112.203901,Chen2017Exceptional,Liu2016Metrology}. 

In this Letter, we use a QAH topological circulator to probe microwave-photon-induced perturbations of EMP transport and examine the response near an exceptional point. Pump--probe measurements reveal a strongly frequency-selective modification of the EMP transmission, including pump frequencies at which the EMP response remains remarkably insensitive to increasing microwave power. We further show that the pump-induced change in transmission magnitude near the EP is approximately twice that observed away from the EP, demonstrating an enhanced microwave response to perturbations in the coupled EMP--resonator system.






\section{Device and Experimental Setup}
The device studied here uses the same three-port QAH circulator architecture reported previously in Ref.~\cite{Martinez2026Circulators}. As shown in Fig.~\ref{Figure1}, the device consists of a $100~\mu\mathrm{m}$-radius Cr-doped $(\mathrm{Bi},\mathrm{Sb})_2\mathrm{Te}_3$ EMP resonator coupled to three lumped-element LC resonators, one at each port. Each LC resonator incorporates a wire-loop inductor with $L=180~\mathrm{nH}$ and a parasitic capacitance $C_{\mathrm{stray}}$, giving a resonance frequency $\omega_{\mathrm{LC}}=1/\sqrt{LC_{\mathrm{stray}}}$ that is designed to lie near the EMP resonance. The EMP resonance frequency depends on the chiral edge-mode dispersion and can be written as\cite{PhysRevLett.77.570}
\begin{equation}
\omega_{\rm EMP}=\left[\frac{\sigma_{xy}}{2\pi\epsilon_0\epsilon_{\mathrm{eff}}}\left(
\ln\frac{2}{|q|w}+1\right)+v_D\right]q
\end{equation}
\noindent where $q$ is the EMP wave vector, $\sigma_{xy}=Ce^2/h$ is the Hall conductivity with a Chern number \(C = 1\), \(\epsilon_{\mathrm{eff}}\) is the effective dielectric constant, $v_D$ is the drift velocity of the one-dimensional edge state, and $w$ is the effective width of the EMP mode. At each port, the EMP resonator is capacitively coupled to the corresponding LC resonator through an edge capacitance $C_{\mathrm{edge}}$ formed between the Au pad and the QAH mesa, as illustrated in Fig.~\ref{Figure1}. Additional details of the device design and operating principle are given in Ref.~\cite{PhysRevResearch.6.013081,Martinez2026Circulators}.

\begin{figure}[t]
\centering
\includegraphics[width=0.95\columnwidth]{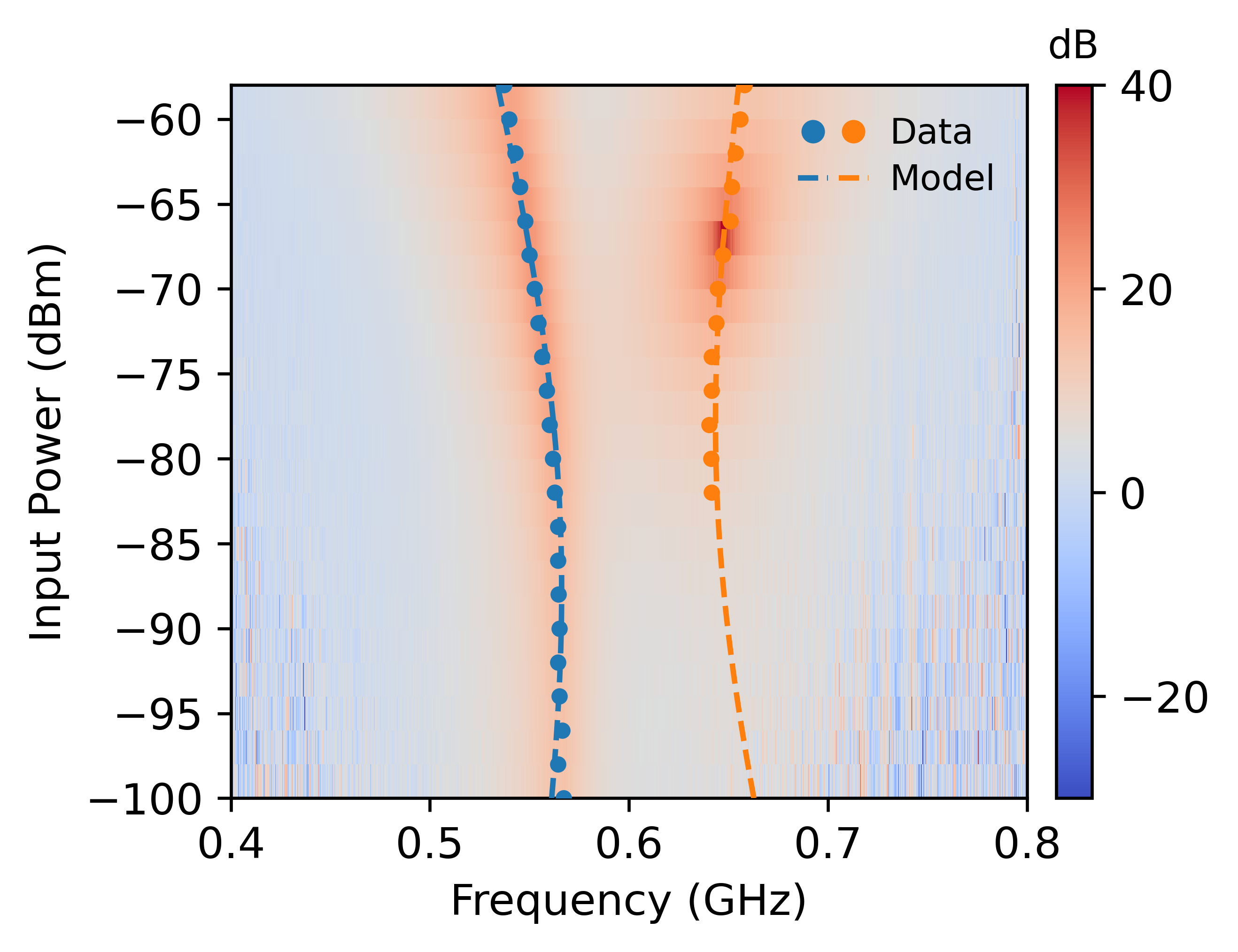}
\caption{{\bf Power-dependent hybridization of the LC and EMP modes.} Color map of the nonreciprocal transmission contrast, \(|S_{31}|-|S_{32}|\), as a function of frequency and input power. The extracted lower- and upper-frequency resonance branches are indicated by blue and orange circles, respectively, while the dashed curves show a fit to the two-mode coupled-resonator model.}\label{Figure3} 
\vspace{-10pt}
\end{figure}

\begin{figure*}[!t]
\centering
\includegraphics[width=.95\textwidth]{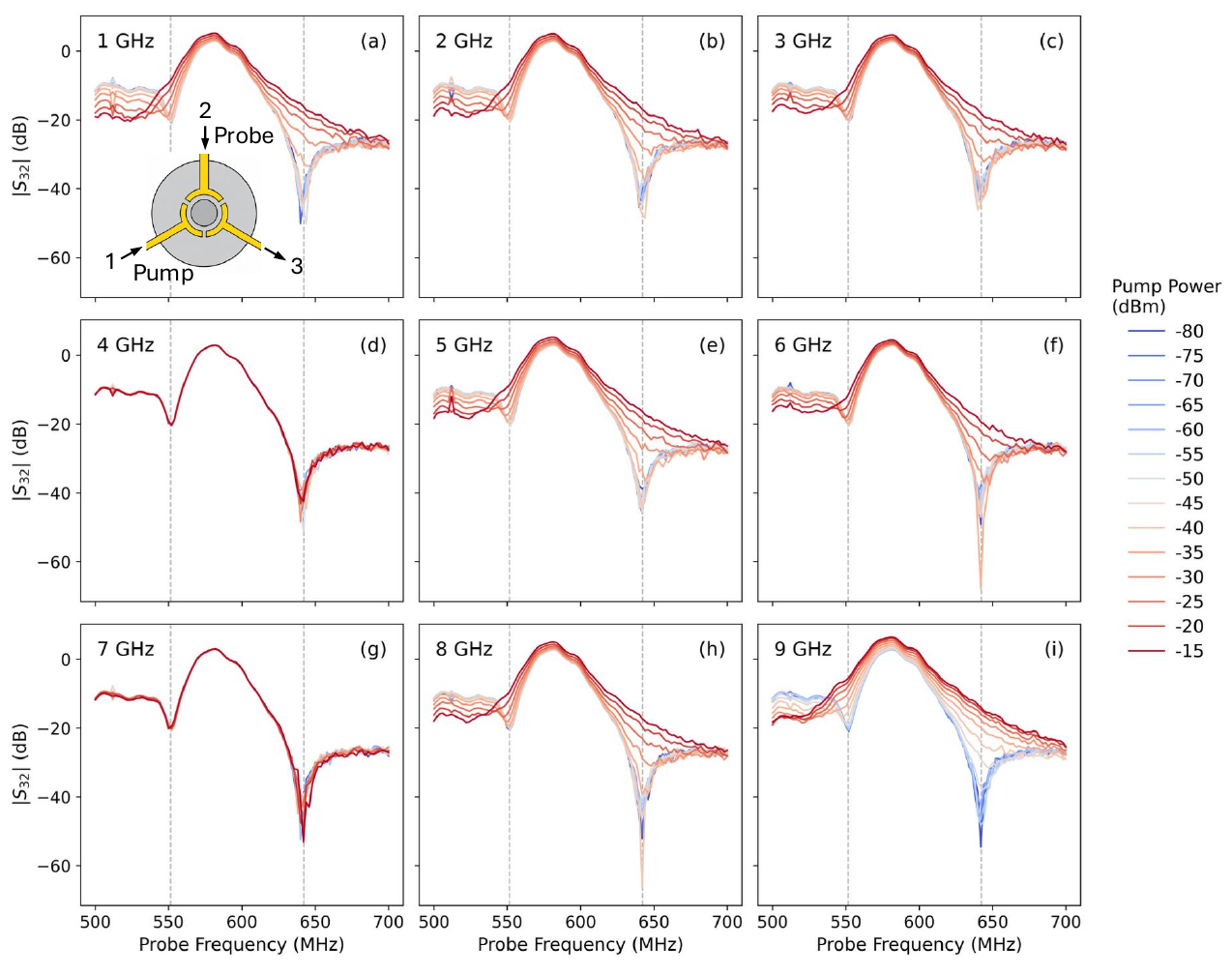}
\caption{\label{Figure4} Microwave transmission spectra between 500 and 700 MHz under various pump powers. A continuous-wave pump signal from the MXG is injected through Port 1, with the pump frequency varied from 1 to 9 GHz over a range of power levels. The transmission spectra are measured using the RFSoC/QICK setup by applying a \(2.2~\mu\mathrm{s}\)-long excitation pulse to Port 2 and measuring the transmitted signal at Port 3, as shown in the inset of (a). The two dashed lines mark the frequencies at \(551.5\) and \(642.3~\mathrm{MHz}\).}
\end{figure*}

The device is characterized at millikelvin temperatures using complementary microwave measurement platforms. The device is mounted at the mixing chamber stage (\(\sim 50\)--\(300~\rm mK\)) of a dry dilution refrigerator. We use both a vector network analyzer (VNA) and a Xilinx Zynq UltraScale+ RFSoC board with QICK FPGA overlay for RF signal generation and acquisition. Details about the fabrication process and experimental setup are provided in Supplementary Section I.

\begin{figure}[!t]
\centering
\includegraphics[width=0.85\columnwidth]{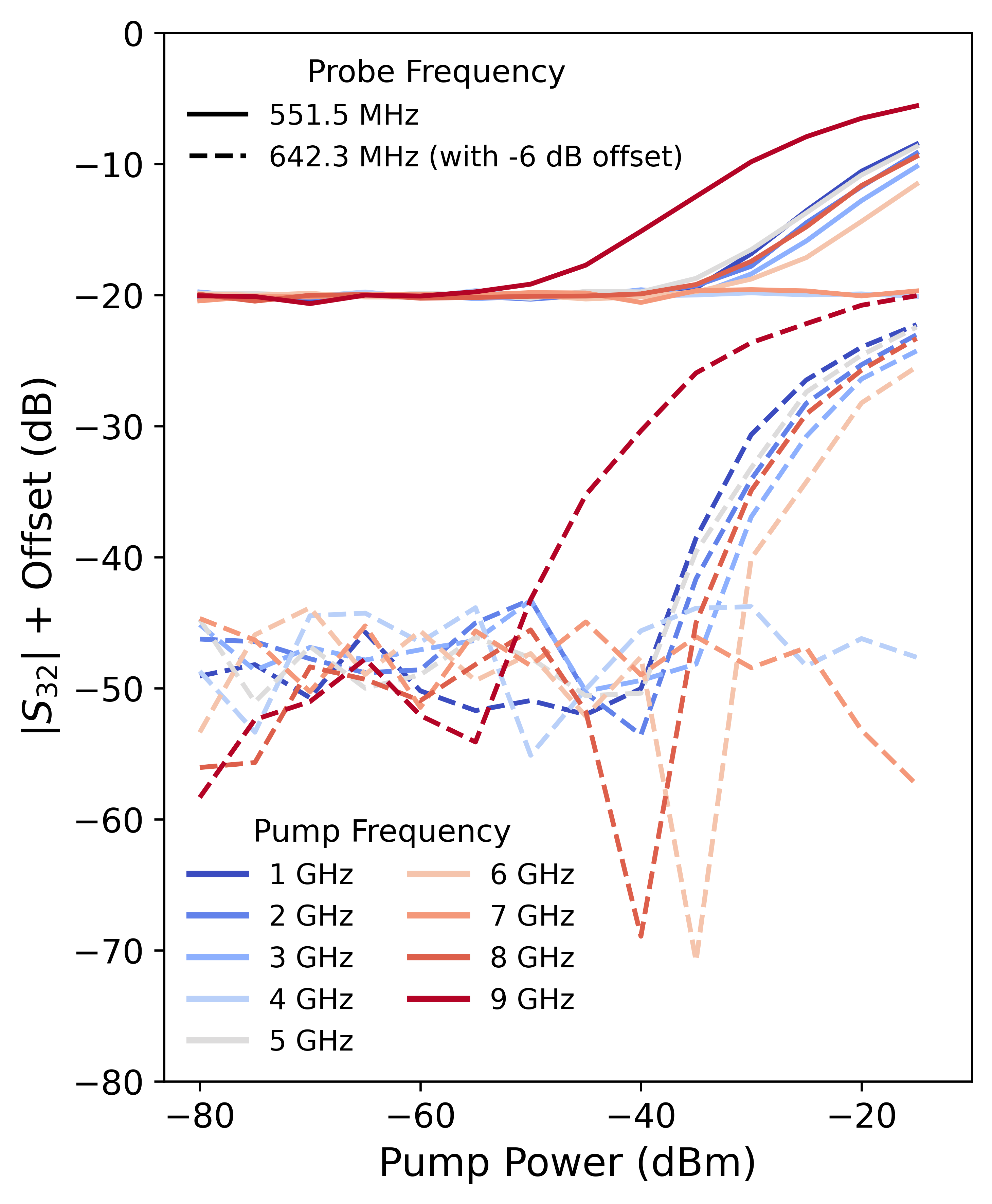}
\caption{\label{Figure5} {\bf Pump-power dependence of the transmission response at two probe frequencies.} Transmission magnitude \(|S_{32}|\) as a function of pump power for pump frequencies from 1 to 9 GHz, measured at probe frequencies of 551.5 MHz (solid lines) and 642.3 MHz (dashed lines). The 642.3 MHz traces are vertically offset by \(-6\) dB for clarity. The response at 642.3 MHz, near the exceptional point, exhibits a substantially stronger pump-power dependence for several pump frequencies than the response at 551.5 MHz, away from the exceptional point, while selected pump frequencies show comparatively weak perturbation.}
\end{figure}

\section{Circulator Transmission Response}
The measured microwave response reveals strong power-dependent nonreciprocity across $0.4$--$0.8~\mathrm{GHz}$, with the transmission contrast exceeding $50~\mathrm{dB}$ near $642.3~\mathrm{MHz}$. Figure 2 shows the microwave transmission response of the device as a function of frequency and input power. Figures 2(a) and 2(b) display the measured \(|S_{31}|\) and \(|S_{32}|\), respectively, while Fig. 2(c) shows the nonreciprocal transmission contrast, \(|S_{31}|-|S_{32}|\). Both transmission channels exhibit a broad resonance centered near \(580~\mathrm{MHz}\).  The nonreciprocity is most pronounced around an input power of \(-68~\mathrm{dBm}\), indicated by the white dashed lines in Figs. 2(a)–2(c). The \(|S_{31}|-|S_{32}|\) 2D map exhibits an avoided-crossing-type splitting as a function of input power as shown in Fig. \ref{Figure2}(c). The corresponding line cuts are shown in Figs. 2(d)–2(f). At this power, \(|S_{32}|\) exhibits a deep minimum near \(642.3~\mathrm{MHz}\). A second, broader nonreciprocal feature with a contrast exceeding \(20~\mathrm{dB}\) is also observed near \(551.5~\mathrm{MHz}\), as shown in Fig. \ref{Figure2}(f). 

This coupled plasmon-photon system can be described using an effective non-Hermitian coupled-mode model (Supplementary Section II). The inequivalent chiral propagation paths around the QAH mesa produce different phase accumulation and dissipation, leading to asymmetric effective coupling between the LC and EMP resonators \cite{Martinez2026Circulators, PhysRevLett.134.153801}. Interference between the EMP and parasitic capacitive transmission pathways consequently gives rise to strongly nonreciprocal amplitude responses. Because the effective directional couplings depend on the propagation and dissipation of the EMP, changes in the underlying QAH transport are directly reflected in the complex microwave scattering response.


We next analyze the avoided-crossing-type splitting shown in Fig.~\ref{Figure3}, which demonstrates strong hybridization between the LC and EMP modes. The two resonance branches are approximately centered near $600~\mathrm{MHz}$, with their minimum separation occurring near $P_0\approx -82~\mathrm{dBm}$. The extracted resonance frequencies are fitted using
\begin{equation}
f_{\pm}(P) = f_{0} + \beta(P-P_{0}) \pm \frac{1}{2} \sqrt{\left[\alpha(P-P_{0})\right]^2+(2g_f)^2},
\label{equation1}
\end{equation}
\noindent where $f_{\pm}(P)$ are the frequencies of the upper and lower hybridized modes, $f_0$ is the resonance frequency at the bare-mode crossing, and $P_0$ is the input power at which the two uncoupled modes are resonant. The parameter $\alpha$ describes the power dependence of the bare-mode detuning, $\beta$ accounts for the common shift of the average resonance frequency with power, and $g_f=g/2\pi$ is the effective coupling strength expressed in frequency units. The fit yields $g_f=39.66\pm1.18~\mathrm{MHz}$. Additional results showing the avoided crossing with reduced coupling strength at higher temperatures are provided in Supplementary Section III.

Although the observed avoided crossing reflects strong mode hybridization, it does not by itself signify an exceptional point. The coupled EMP--LC system is intrinsically non-Hermitian because both modes experience dissipation, while the chiral EMP propagation produces direction-dependent phase accumulation and asymmetric effective couplings between the resonators. Consequently, the eigenfrequencies are generally complex and the two directional coupling amplitudes need not be equal. In an ordinary avoided crossing, finite coupling maintains a nonzero separation between the two hybridized modes. In contrast, as one directional effective coupling approaches zero while the reverse coupling remains finite, the complex eigenvalues and corresponding eigenvectors coalesce, driving the reduced coupled-mode system toward a Hatano–Nelson-type exceptional point.

The upper hybridized branch is particularly relevant to the sensing measurements discussed below. Near $642.3~\mathrm{MHz}$, $|S_{32}|$ exhibits a pronounced suppression and a correspondingly strong nonreciprocal response, indicating operation in the regime where the directional coupling is strongly modified and the system approaches the EP. We therefore use the transmission response at $642.3~\mathrm{MHz}$ as the near-EP probe, while the response at $551.5~\mathrm{MHz}$ provides a reference farther from the EP. This comparison enables the enhanced response to microwave-photon-induced perturbations near the EP to be quantified directly. 

\section{Probing the Microwave Photon-Induced Perturbation}
To investigate the response of EMP transport to microwave-photon perturbations, we perform pump--probe measurements in which a continuous-wave microwave pump is applied to Port 1 while the transmission from Port 2 to Port 3 is measured using the RFSoC/QICK setup. The pump frequency is varied from $1$ to $9~\mathrm{GHz}$ over a broad range of pump powers, while the probe frequency is swept from $500$ to $700~\mathrm{MHz}$. The sample temperature is maintained at a constant value while the pump power is swept. As shown in Fig.~\ref{Figure4}, the microwave pump produces a strongly frequency-dependent modification of the EMP transmission spectrum. The effect is particularly pronounced around the transmission feature near $642.3~\mathrm{MHz}$. For several pump frequencies, the spectra remain relatively unchanged at low pump powers and begin to evolve more strongly as the pump power is increased, with the most prominent changes occurring approximately above $-50$ to $-40~\mathrm{dBm}$.

The magnitude and character of this response vary substantially with pump frequency. For example, at $9~\mathrm{GHz}$ the transmission near $642.3~\mathrm{MHz}$ changes by approximately $30~\mathrm{dB}$ between the low- and high-power limits. At $6~\mathrm{GHz}$, an even deeper suppression develops near $642.3~\mathrm{MHz}$ at intermediate pump powers, demonstrating that the response can also be nonmonotonic with pump power. Comparable power-dependent modifications are observed for several other pump frequencies, including $1$--$3$, $5$, and $8~\mathrm{GHz}$. Additional pump-power-dependent measurements are presented in Fig.~\ref{FigureS3} and discussed in the Supplementary Section IV. In striking contrast, at pump frequencies of $4$ and $7~\mathrm{GHz}$ [Figs.~\ref{Figure4}(d) and \ref{Figure4}(g)], the transmission spectra remain nearly unchanged over the entire measured pump--power range. Within our experimental sensitivity, microwave irradiation at these two frequencies therefore produces little detectable perturbation of the EMP transmission.

Importantly, the insensitive response at $4$ and $7~\mathrm{GHz}$ does not follow a simple monotonic trend with pump frequency. These frequencies are interspersed between neighboring drive frequencies that
produce substantial changes in the same EMP transmission feature. Thus, the observed response cannot be explained simply by a gradual increase or decrease in microwave coupling with drive frequency. Instead, the data demonstrate a highly frequency-selective interaction between the applied microwave field and the EMP system.

This behavior is qualitatively distinct from the microwave-induced breakdown of QAH transport reported in Ref.~\cite{Roper2025Breakdown}. In that work, microwave irradiation between $1$ and $25~\mathrm{GHz}$ enhanced longitudinal bulk conductance above a frequency-dependent threshold, with the breakdown attributed primarily to RF-induced heating of electron--hole puddles and the resulting enhancement of hopping transport. The threshold microwave amplitude was found to decrease systematically with increasing drive frequency. In contrast, our measurements show that increasing microwave power can produce a strong EMP response at one pump frequency but essentially no detectable modification at a nearby frequency. The observed perturbation therefore cannot be described by a simple frequency-monotonic heating picture alone.

The existence of such \textit{insensitive} pump frequencies suggests
that the response is governed by the frequency-dependent coupling of
the applied microwave field to the QAH device and its electromagnetic
environment. Possible contributions include frequency-dependent
microwave absorption, coupling to localized charge degrees of freedom,
and the impedance of the coupled EMP--resonator structure. \cite{Enkner2025Tunable} Although the microscopic origin of this selectivity requires further investigation, these measurements demonstrate that EMP transport can remain remarkably robust against microwave irradiation at selected drive frequencies while exhibiting a strong response at neighboring frequencies.


To evaluate whether the EP enhances the response to microwave-photon-induced perturbations, we compare the pump-power
dependence of the transmission magnitude at two characteristic probe
frequencies. Figure~\ref{Figure5} shows $|S_{32}|$ at $551.5~\mathrm{MHz}$, corresponding to a resonant response away from
the exceptional point, and at $642.3~\mathrm{MHz}$, near the EP.
For clarity, the $642.3~\mathrm{MHz}$ traces are vertically offset by $-6~\mathrm{dB}$. At low pump powers, the transmission at both probe
frequencies is nearly independent of the applied microwave perturbation. As the pump power increases, however, the response near the EP exhibits a substantially larger change in transmission magnitude.

We quantify the perturbation response by the change in $|S_{32}|$ expressed in dB over the same pump-power interval, relative to its low-power value. Using this measure, the variation in $|S_{32}|$ near $642.3~\mathrm{MHz}$ is approximately twice as large as that observed at $551.5~\mathrm{MHz}$. Thus, the enhancement refers to the magnitude of the transmission response in dB, rather than to a
factor-of-two increase in the linear transmission amplitude. The larger dB excursion near $642.3~\mathrm{MHz}$ indicates an enhanced susceptibility of the microwave scattering response to the applied perturbation in the vicinity of the EP.

This observation is consistent with the increased susceptibility to perturbations expected near an exceptional point, in which the coalescence of eigenmodes and eigenvalues in a non-Hermitian system can produce an
enhanced response to perturbations.\cite{PhysRevLett.112.203901, Lau2018} Conventional EP sensing studies often quantify this enhancement through changes in eigenfrequency splitting.\cite{Hodaei2017HigherOrderEP} In the present experiment, however, we do not use the eigenfrequency splitting itself as the sensing observable. Instead,
the experimentally accessible quantity is the pump-induced change in the scattering amplitude $|S_{32}|$. The approximately twofold larger dB response near the EP therefore demonstrates an enhancement of the
microwave transmission response of the coupled EMP--resonator system. We note that this enhancement refers specifically to the measured transmission response; establishing an improvement in noise-limited sensing precision would additionally require a quantitative analysis of the signal-to-noise ratio and measurement noise.

\section{Conclusion}
In conclusion, we investigated the microwave response of a QAH-based topological circulator coupled to lumped-element resonators. The device exhibits strong nonreciprocal transmission and a pronounced power-dependent response associated with the coupled EMP--resonator modes. Pump--probe measurements reveal a highly frequency-selective microwave-photon-induced perturbation, with the EMP transmission remaining remarkably robust at selected pump frequencies, including $4$ and $7~\mathrm{GHz}$. In addition, the transmission response near the exceptional point is approximately twice as large as that observed away from the EP. These results highlight the interplay between topological edge transport, non-Hermitian mode hybridization, and microwave perturbations, and reveal a range of emergent phenomena, including strongly frequency-selective responses and EP behavior, enabled by the coupled topological-circulator device structure.

We would like to thank Kristi Beck, Bevin Huang, Sayan Patra, and Vito Iaia for helpful discussions. This work was performed under the auspices of the US Department of Energy by Lawrence Livermore National Laboratory under Contract No. DE-AC52-07NA27344. This work was supported by the Laboratory Directed Research and Development (LDRD) Program at Lawrence Livermore National Laboratory (LLNL) under projects 26-ERD-010, and supported in part by the U.S. Department of Energy (DOE) Office of Science, Office of High Energy Physics, under Work Proposal Numbers SCW1915 and SCW1508 to LLNL. The authors also acknowledge support from NSF Convergence Accelerator program under Grant 2040737.

\bibliography{Reference}

\clearpage
\onecolumngrid
%
%
%
%
%
%
%



\setcounter{equation}{0}
\renewcommand{\theequation}{S\arabic{equation}}

\setcounter{figure}{0} 
\renewcommand{\thefigure}{S\arabic{figure}} 
\renewcommand{\theHfigure}{supplement.figure.\arabic{figure}}

\setcounter{section}{0}
\renewcommand{\thesection}{\Roman{section}}
\makeatletter
\def\@email#1#2{%
 \endgroup
 \patchcmd{\titleblock@produce}
  {\frontmatter@RRAPformat}
  {\frontmatter@RRAPformat{\produce@RRAP{*#1\href{mailto:#2}{#2}}}\frontmatter@RRAPformat}
  {}{}
}%
\makeatother

\preprint{AIP/123-QED}

\title[]{Supplementary Information: Probing the Quantum Anomalous Hall Transport Under Microwave Irradiation Using a Topological Circulator}

\noindent
{\raggedright
\sffamily
\bfseries
\fontsize{14.4pt}{17pt}\selectfont
Supplementary Material for\\[0.25em]
``Probing Quantum Anomalous Hall Transport Under Microwave Irradiation Using a Topological Circulator''
\par}


\section{\label{sec:level1}Device Fabrication and Experimental Setup}

The six-quintuple-layer \(\mathrm{Cr}_{0.12}(\mathrm{Bi}_{0.26}\mathrm{Sb}_{0.62})_2\mathrm{Te}_3\) thin films were grown by ultrahigh-vacuum molecular beam epitaxy. Epi-ready semi-insulating GaAs (111)B substrates (\(\rho > 10^6\,\Omega\mathrm{cm}\)) were pre-annealed in the growth chamber at temperatures up to \(600\,^\circ\mathrm{C}\) under Te-rich conditions. During co-evaporation, high-purity Bi (99.9999\%), Te (99.9999\%), and Cr (99.99\%) were supplied from Knudsen effusion cells, while Sb (99.999\%) was supplied from a cracker cell. Following growth, the circulator devices were fabricated using standard photolithography. The mesa disk was first defined by photolithography and etched by reactive-ion etching using a \(\mathrm{CF}_4/\mathrm{Ar}\) plasma. The coplanar waveguides and ground planes were then patterned by photolithography, followed by electron-beam evaporation of 10 nm Ti and 100 nm Au and a subsequent lift-off process.

We characterized the circulator in a dry dilution refrigerator (Bluefors BF-LD400). The input lines pass through a series of cryogenic attenuators providing 20 dB, 10 dB, and 10 dB of attenuation for the 3K, still, and intermediate cold flanges, respectively. The output line includes 15 dB of isolation (QuinStar QCI-M6009001AM) and is connected to a 40 dB high-electron-mobility transistor (HEMT) amplifier (LNF-LNC0.3-14A) through a NbTi superconducting RF cable. The sample is mounted inside a dual-cylinder magnetic shield attached to the mixing-chamber flange and oriented upward toward the still flange (1 K stage). This configuration cools the circulator to approximately 50 mK because of its proximity to the still stage, while the mixing-chamber flange remains below 10 mK. The details of the experimental configuration used for the vector-network-analyzer measurements of the microwave transmission spectra can be found in Ref.~\cite{Martinez2026Circulators}. In addition, we used a Zynq UltraScale+ RFSoC ZCU111 board running the QICK (Quantum Instrumentation Control Kit) FPGA overlay to generate and acquire coherent RF signals at frequencies up to 1 GHz \cite{stefanazzi2022qick}. A tunable microwave attenuator (minicircuits, RCDAT-6000-90) and a room temperature amplifier (Mini-Circuits, ZX60-P33ULN+) are employed to utilize the full DAC/ADC range.

\section{Coupling strength and its dependence on the input power}
For coupled LC and edge magnetoplasmon (EMP) resonators, the coupled-mode Hamiltonian can be written as
\begin{equation}
\frac{H_{\rm eff}}{\hbar}
=
\begin{pmatrix}
\omega_{\rm LC}-i\kappa_{\rm LC}/2 & G_{12} \\
G_{21} & \omega_{\rm EMP}-i\kappa_{\rm EMP}/2
\end{pmatrix}.
\end{equation}
\noindent where \(\omega_{\rm LC}\) is the angular resonance frequency of the LC mode, \(\omega_{\rm EMP}\) is the angular resonance frequency of the EMP mode, \(G_{12}\) is the coupling from the EMP mode into the LC mode, \(G_{21}\) is the coupling from the LC mode into the EMP mode, \((2g_f)^2=\frac{4 G_{12}G_{21}}{(2\pi)^2}\) is the effective coupling strength, and \(\kappa_{\rm LC}\) and \(\kappa_{\rm EMP}\) are the energy-decay rates of the LC and plasmonic modes, respectively. Approximately, \(\kappa_{\rm LC} \approx \frac{\omega_{\rm LC}}{Q_{\rm LC}}=2\pi \Delta f_{\rm LC}\) and \(\kappa_{\rm EMP} \approx \frac{\omega_{\rm EMP}}{Q_{\rm EMP}} = 2\pi \Delta f_{\rm EMP}\). The hybridized frequencies, \(f_{\pm}(P)\), are the eigenvalues of this matrix assuming that their uncoupled frequencies vary approximately linearly with power \(P\). The  eigenfrequencies of the coupled system are therefore
\begin{equation}
\omega_{\pm}
=
\frac{\omega_{\rm LC}+\omega_{\rm EMP}}{2}
\pm
\frac{1}{2}
\sqrt{
\left(\omega_{\rm LC}-\omega_{\rm EMP}\right)^2
+
4G_{12}G_{21}
}.
\end{equation}

So the coupled-mode frequencies \(f_{\pm}= \omega_{\pm}/2\pi \) are

\begin{equation}
f_{\pm}=\bar{f}\pm\frac{1}{2}\sqrt{\Delta f^2+(2g_f)^2},
\end{equation}
\noindent where \(\bar{f} =\frac{f_{\rm LC}+f_{\rm EMP}}{2}\) and \(\Delta f = f_{\rm LC}-f_{\rm EMP}\). Near the avoided crossing at $P=P_0$, the two uncoupled mode
frequencies can be approximated as linear functions of power: \(f_{\rm LC}(P)=f_0+\left(\beta+\frac{\alpha}{2}\right)(P-P_0) \), and \(f_{\rm EMP}(P)=f_0+\left(\beta-\frac{\alpha}{2}\right)(P-P_0)\). Their average frequency is therefore \(\frac{f_{\rm LC}(P)+f_{\rm EMP}(P)}{2}=f_0+\beta(P-P_0)\), while their detuning is \(f_{\rm LC}(P)-f_{\rm EMP}(P) =\alpha(P-P_0)\). Substituting these expressions into the coupled-mode eigenfrequencies yields Eq. (2) in the main text
\begin{equation}
f_{\pm}(P)=f_0+\beta(P-P_0)\pm\frac{1}{2}
\sqrt{\left[\alpha(P-P_0)\right]^2+(2g_f)^2}.
\end{equation}
\noindent where \(f_0\) is the bare-mode crossing frequency, \(P_0\) is the power at the bare-mode crossing, \(\alpha\) is the rate at which the bare-mode detuning changes with \(P\), \(\beta\) is the average frequency drift of the two bare modes with \(P\). At \(P=P_0\), \(f_{\pm}(P_0) = f_0\pm g_f\), so the minimum mode splitting is \(f_{+}-f_{-}=2g_f\).

At a circulator temperature of \(T \sim 50\, \rm mK\), the fitting parameters are \(f_0 = 604.83\ \rm MHz\), \(P_0 = -82.812 \ \rm {dBm}\), \(\alpha = 0.003694 \) GHz/dBm, \(\beta = -0.000414 \ \rm{GHz/dBm}\), \( g_f=g/2\pi = 39.66 \ \rm{MHz}\), and \(
2g/2\pi   = 79.32\ \rm {MHz}\). To characterize the coupling strength between the LC and EMP resonators, we consider a commonly used criterion for identifying the strong-coupling regime, given by ~\cite{BarraBurillo2021}
\begin{equation}
g>\frac{\kappa_{\rm LC}+\kappa_{\rm EMP}}{4},
\end{equation}

The coupling strength between the LC and EMP resonators can be further quantified by the cooperativity,~\cite{Wang2019} which characterizes the coherent coupling relative to the dissipation rates of the two resonators,
\begin{equation}
C=\frac{4g^2}{\kappa_{\rm LC}\kappa_{\rm EMP}},
\end{equation}
with \(C>1\) indicating that the coherent coupling is sufficiently strong relative to the dissipative rates to produce significant mode hybridization.

Across the full \(500\)–\(600\) MHz frequency range, the approximate strong-coupling condition is
\(\frac{g}{2\pi} \gtrsim 15\text{--}18 \ \mathrm{MHz}\) as shown in Table \ref{tab:strong_coupling}. For example,  at \(550\,\mathrm{MHz}\), the linewidths of the LC and EMP modes are \(\Delta f_{\rm LC} = \frac{550}{24} \approx 22.9\ \mathrm{MHz}\) and \( \Delta f_{\rm EMP} = \frac{550}{13} \approx 42.3\ \mathrm{MHz}\), respectively. Therefore, the strong-coupling threshold is \(g_f > \frac{\Delta f_{\rm LC}+\Delta f_{\rm EMP}}{4}=16.3\ \mathrm{MHz}\). 

\begin{table}[ht]
\centering
\caption{Estimated linewidths and strong-coupling thresholds for the LC and EMP modes.}
\label{tab:strong_coupling}
\setlength{\tabcolsep}{10pt} 
\begin{tabular}{cccc}
\hline
Resonance frequency
&
$\Delta f_{\rm LC}$, $Q_{\rm LC}=24$
&
$\Delta f_{\rm EMP}$, $Q_{\rm EMP}=13$
&
Strong-coupling threshold $g_f$
\\
\hline
$500\,\mathrm{MHz}$
&
$20.8\,\mathrm{MHz}$
&
$38.5\,\mathrm{MHz}$
&
$14.8\,\mathrm{MHz}$
\\

$550\,\mathrm{MHz}$
&
$22.9\,\mathrm{MHz}$
&
$42.3\,\mathrm{MHz}$
&
$16.3\,\mathrm{MHz}$
\\

$600\,\mathrm{MHz}$
&
$25.0\,\mathrm{MHz}$
&
$46.2\,\mathrm{MHz}$
&
$17.8\,\mathrm{MHz}$
\\
\hline
\end{tabular}
\end{table}
Thus, the LC resonance is relatively broad, but it is not necessarily too broad to exhibit strong coupling. The experimentally extracted coupling strength, \(g_f = 39.66 \pm 1.18\ \mathrm{MHz}\) at \(50\ \mathrm{mK}\) is more than twice this threshold. Equivalently, \(C \approx 6 > 1 \) at \(550\ \mathrm{MHz}\). Therefore, the system reaches the strong LC-EMP coupling regime, implying a strong coupling between the microwave photon and the edge magnetoplasmons.

\begin{figure}[t]
\centering
\includegraphics[width=0.95\columnwidth]{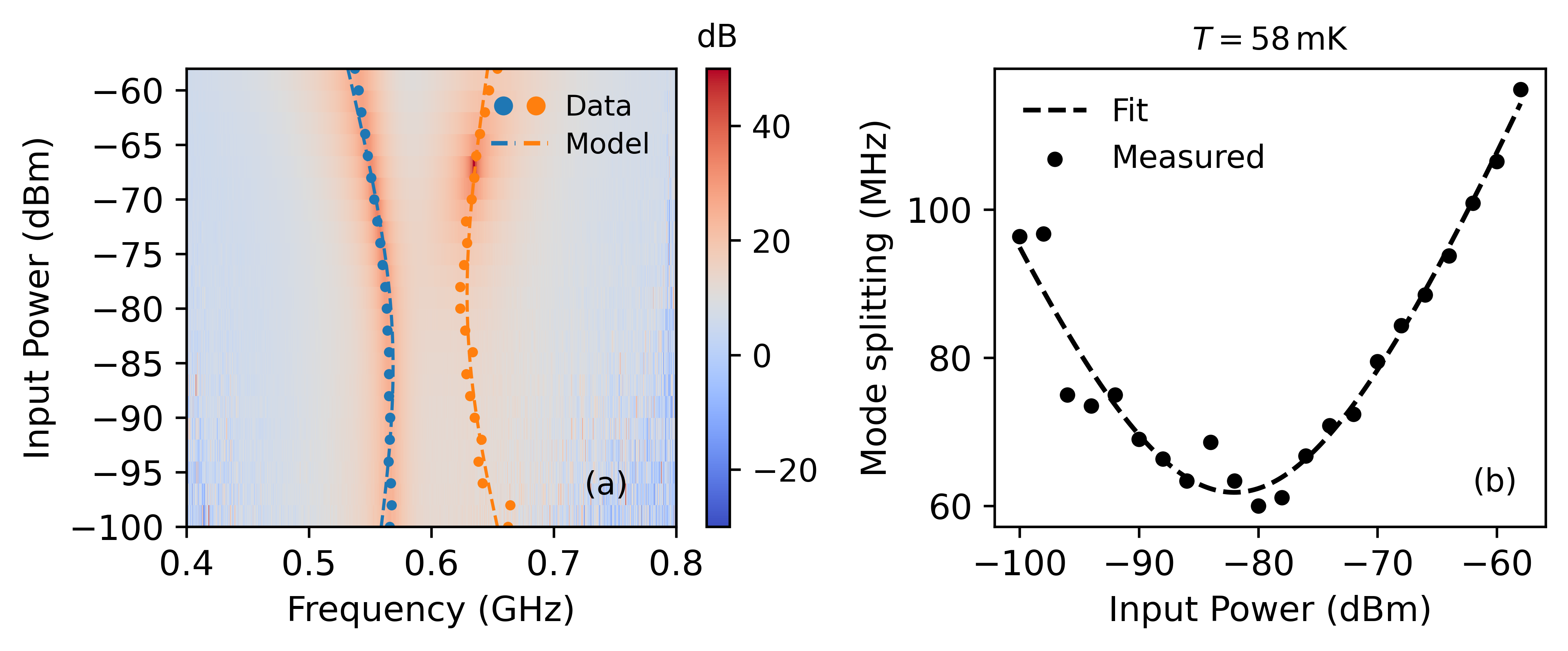}
\caption{{\bf Power-dependent hybridization of the LC and EMP modes at \(T = 58\ \mathrm{mK}\).} (a) Color map of the nonreciprocal transmission contrast, \(|S_{31}|-|S_{32}|\), as a function of frequency and input power. The blue and orange circles mark the extracted lower- and upper-frequency hybridized branches, respectively, and the dashed curves represent fits to the two-mode coupled-resonator model. (b) Power-dependent mode splitting determined from the separation between the two extracted branches. The circles show the experimental splitting, while the dashed curve indicates the corresponding coupled-mode fit.
}\label{FigureS1} 
\end{figure}

\section{Circulator Transmission Spectra at a slightly Higher Temperature}

\begin{figure}[t]
\centering
\includegraphics[width=0.95\columnwidth]{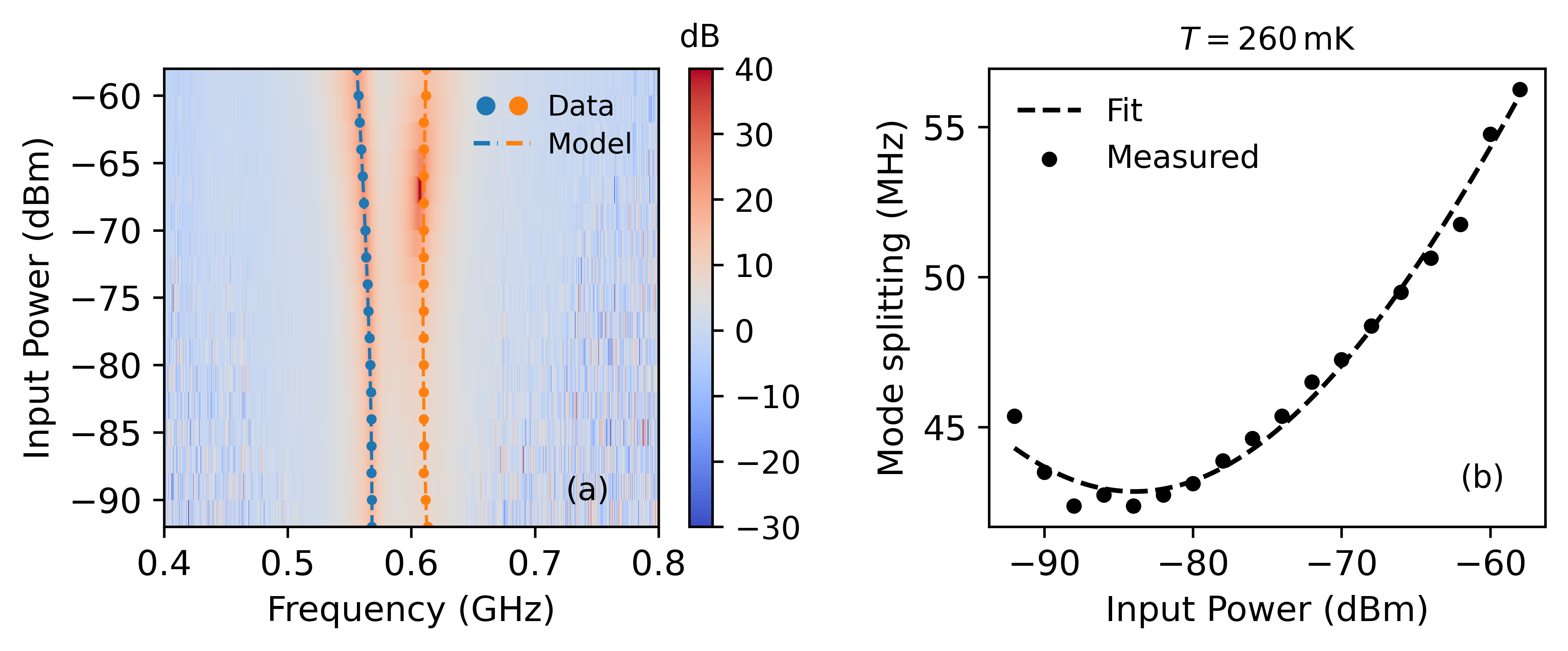}
\caption{{\bf Power-dependent hybridization of the LC and EMP modes at \(T = 260\ \mathrm {mK}\).} (a) Color map of the nonreciprocal transmission contrast, \(|S_{31}|-|S_{32}|\), as a function of frequency and input power. The blue and orange circles mark the extracted lower- and upper-frequency hybridized branches, respectively, and the dashed curves represent fits to the two-mode coupled-resonator model. (b) Power-dependent mode splitting determined from the separation between the two extracted branches. The circles show the experimental splitting, while the dashed curve indicates the corresponding coupled-mode fit.
}\label{FigureS2} 
\end{figure}

We observed that the LC--EMP coupling strength decreases with increasing temperature. At \(T \sim 58\,\mathrm{mK}\), the coupling strength is reduced to \(g_f = 30.92 \pm 1.11\,\mathrm{MHz}\), as extracted from the fit shown in Fig.~\ref{FigureS1}. The fitting parameters are \(f_0 = 599.097\, \rm MHz\), \(P_0 = -82.015\, \rm {dBm}\), \(\alpha = 0.004004\, \rm GHz/dBm\), and \(\beta = -0.000420 \, \rm{GHz/dBm}\). Further increasing the temperature to \(T \sim 260\,\mathrm{mK}\), the coupling strength is further reduced to \(g_f = 21.43 \pm 0.10\,\mathrm{MHz}\), as extracted from the fit shown in Fig.~\ref{FigureS2}. The fitting parameters are \(f_0 = 588.84\, \rm MHz\), \(P_0 = -83.936\, \rm {dBm}\), \(\alpha = 0.001393\, \rm GHz/dBm\), and \(\beta = -0.000188\, \rm{GHz/dBm}\).

\section{Circulator Transmission Spectra at Additional Pump Frequencies}

\begin{figure*}[!t]
\centering
\includegraphics[width=.95\textwidth]{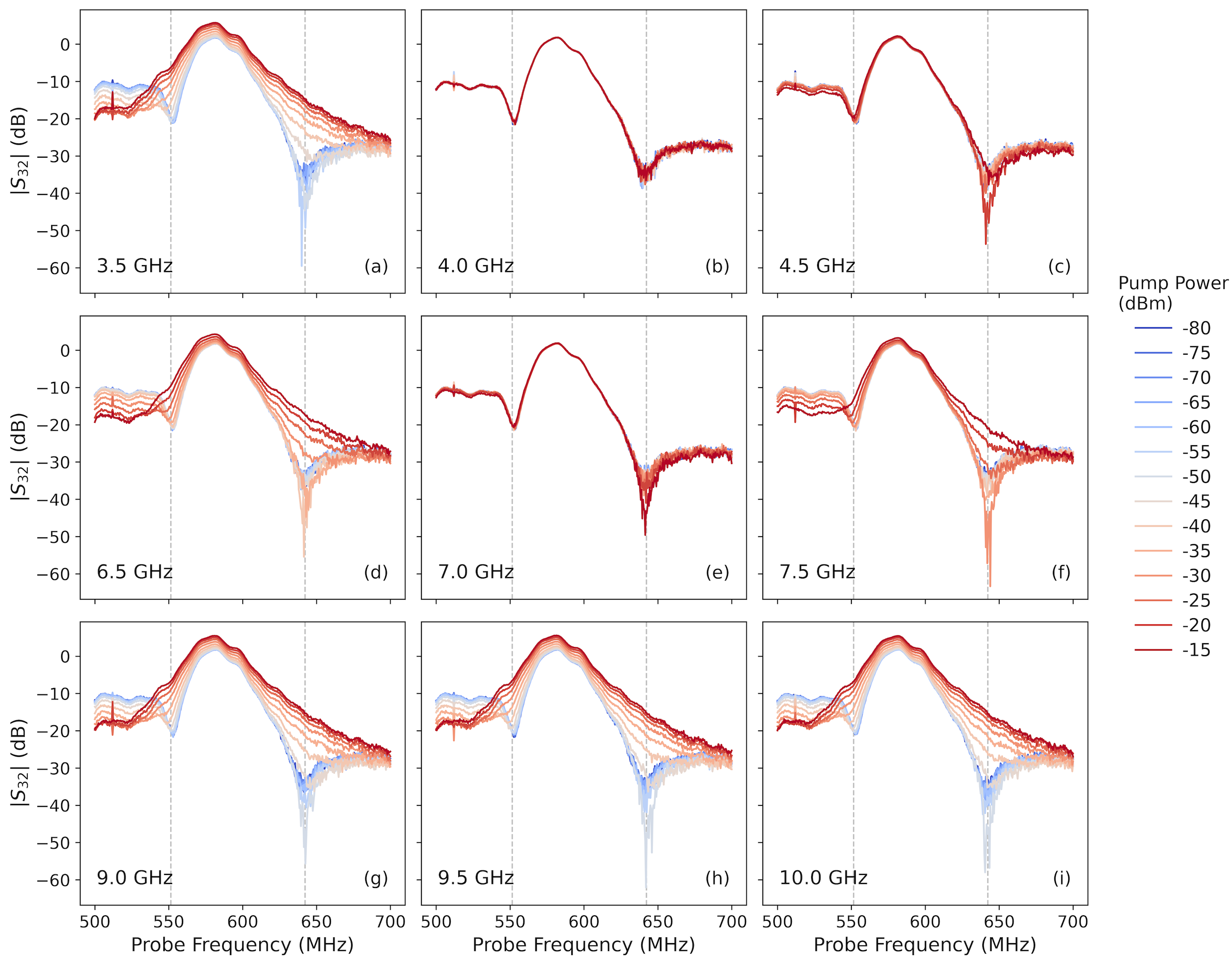}
\caption{\textbf{Extended pump-frequency-dependent transmission measurements corresponding to Fig.~\ref{Figure4} of the main text.} Microwave transmission spectra between 500 and 700 MHz under various pump powers. A continuous-wave pump signal is injected through Port 1, with the pump frequency varied at 3.5, 4.0, 4.5, 6.5, 7.0, 7.5, 9.0, 9.5, and 10.0 GHz over a range of power levels.}\label{FigureS3}
\end{figure*}

To further examine the frequency dependence of the microwave-induced perturbation, we measured the transmission spectra at additional pump frequencies of $3.5$, $4.0$, $4.5$, $6.5$, $7.0$, $7.5$, $9.0$, $9.5$, and $10.0~\mathrm{GHz}$, as shown in Fig.~\ref{FigureS3}. The pump power is varied from approximately $-80$ to $-15~\mathrm{dBm}$, while the probe frequency is swept from $500$ to $700~\mathrm{MHz}$. The two dashed vertical lines mark the characteristic probe frequencies at $551.5$ and $642.3~\mathrm{MHz}$.

The additional measurements confirm that the perturbation of the EMP transmission is strongly dependent on pump frequency. At $3.5~\mathrm{GHz}$, the transmission near $642.3~\mathrm{MHz}$ evolves substantially with increasing pump power, accompanied by changes in the broader spectral response. In contrast, the spectra at $4.0~\mathrm{GHz}$ remain nearly overlapped over the entire pump-power range, whereas a pronounced power-dependent modification reappears at $4.5~\mathrm{GHz}$. A similar behavior is observed around $7~\mathrm{GHz}$: the spectra at $6.5$ and $7.5~\mathrm{GHz}$ exhibit strong pump-power-dependent changes, particularly near $642.3~\mathrm{MHz}$, while the response at $7.0~\mathrm{GHz}$ is comparatively insensitive to pump power.

At higher pump frequencies of $9.0$, $9.5$, and
$10.0~\mathrm{GHz}$, the transmission again shows pronounced power-dependent evolution. The feature near $642.3~\mathrm{MHz}$ undergoes particularly large changes in magnitude and lineshape as the
pump power is increased, while the feature near $551.5~\mathrm{MHz}$ also exhibits a measurable but generally weaker response. These measurements reinforce that the microwave-induced perturbation does not
vary monotonically with pump frequency. Instead, strongly responsive pump frequencies are interspersed with comparatively insensitive frequencies, most notably near $4$ and $7~\mathrm{GHz}$, further
supporting a frequency-selective coupling between the applied microwave field and the EMP--resonator system.

%

\end{document}